\documentclass[10pt,final,journal,twocolumn]{IEEEtran}

\ifCLASSINFOpdf

\else

\fi

\usepackage{cite}
\usepackage[cmex10]{amsmath}
\usepackage{amsmath}
\usepackage{algorithmic}
\usepackage{stfloats}
\usepackage{multicol,multienum}
\usepackage{multirow}
\usepackage[dvips]{graphicx}

\usepackage[tight,footnotesize]{subfigure}
\usepackage{wrapfig}
\usepackage{color}

 \usepackage{booktabs}
 \usepackage{multirow}

\usepackage{mathtools}
\usepackage{color}
\usepackage{float}
\usepackage{bm}

\usepackage{pifont}

\usepackage[ruled,linesnumbered]{algorithm2e}

\usepackage{bigfoot}

\begin{document}

\title{Embodied AI-empowered Low Altitude Economy: Integrated Sensing, Communications, Computation, and Control (ISC3)}

\author{Yaoqi~Yang,
Yong~Chen,
Jiacheng~Wang,
Geng~Sun,
and Dusit~Niyato,~\IEEEmembership{Fellow,~IEEE}

\thanks{Manuscript received xxx.}

\thanks{Yaoqi~Yang is with the College of Communications Engineering, Army Engineering University of PLA, Nanjing 210000, China. (e-mail:yaoqi$\_$yang@yeah.net)}

\thanks{Yong~Chen is with the Chengdu Fluid Dynamics Innovation Center, Chengdu 610031, China. (e-mail: literature$\_$chen@nudt.edu.cn)}


\thanks{Jiacheng Wang and Dusit Niyato are with the College of Computing and Data Science, NTU, Singapore. (email: jiacheng.wang@ntu.edu.sg; dniyato@ntu.edu.sg)}

\thanks{Geng Sun is with the College of Computer Science and Technology, Jilin University, Changchun 130012, China, and also with the College of Computing and Data Science, Nanyang Technological University, Singapore 639798 (email: sungeng@jlu.edu.cn).}



}

\maketitle

\begin{abstract}
Low altitude economy (LAE) holds immense potential to drive urban development across various sectors. However, LAE also faces challenges in data collection and processing efficiency, flight control precision, and network performance. The challenges could be solved by realizing an integration of sensing, communications, computation, and control (ISC3) for LAE. In this regard, embodied artificial intelligence (EAI), with its unique perception, planning, and decision-making capabilities, offers a promising solution to realize ISC3. Specifically, this paper investigates an application of EAI into ISC3 to support LAE, exploring potential research focuses, solutions, and case study. We begin by outlining rationales and benefits of introducing EAI into LAE, followed by reviewing research directions and solutions for EAI in ISC3. We then propose a framework of an EAI-enabled ISC3 for LAE. The framework's effectiveness is evaluated through a case study of express delivery utilizing an EAI-enabled UAV. Finally, we discuss several future research directions for advancing EAI-enabled LAE.
\end{abstract}


\IEEEpeerreviewmaketitle

\section{Introduction}

With the rapid advancement of aerial vehicle technologies, Low Altitude Economy (LAE) has emerged as a promising area of development. It encompasses the airspace below 1,000 meters, potentially extending to 3,000 meters depending on specific needs\footnote{https://english.news.cn/20240812/cf309c89db024c5bac8b0d70fd9d56b9\\/c.html}. LAE primarily focuses on civil manned and unmanned aerial vehicles (UAVs), encompassing a wide range of activities including passenger transport, cargo delivery, and related services \cite{xianjiang2024creating}. While LAE networks to support efficient, reliable, and robust data communications are essentially an evolution of traditional UAV networks, there are several new challenges to overcome:
\begin{enumerate}
\item \emph{Massive connectivities pose data processing challenges}. The LAE scenario necessitates massive, integrated air-ground connectivity to support numerous UAVs and diverse applications concurrently. A single city may see thousands of UAVs, ranging from air taxis, public service monitoring, to package delivery drones, creating an unprecedented volume of data unlike anything seen in current UAV networks. This poses significant challenges in communication and computation within the LAE context, requiring new communication and data processing paradigms.
\item \emph{Multiple functions drive heterogeneous service networks}. LAE involves a wide variety of services, allowing a single UAV to execute multiple tasks simultaneously, such as data collection, package delivery, and ground traffic monitoring. Moreover, multiple UAVs can collaborate to complete the same task. This differs from current UAV networks, requiring LAE to consider heterogeneous service issues. As a result, such a demand further faces control challenges to LAE, leading to more complex UAVs task scheduling and flying control to maintain safety and security.
\item \emph{Economic impacts demand multiple indicator optimization}. The LAE¡¯s economic viability is directly linked to quality of service (QoS) provided by UAVs. To maximize economic impacts, LAE requires optimized performance indicators, including larger throughput, reduced Age of Information (AoI), improved sensing accuracy, and high data processing efficiency, e.g., for advanced discriminative AI and generative AI algorithms. In comparison to traditional UAV networks, LAE necessitates more efficient and lightweight network optimization approaches to achieve these goals. 
\end{enumerate}

To address the challenges outlined above, several 
traditional approaches might be used for LAE. For example, mobile coded computing \cite{li2020coded} can be employed for data processing, convex optimization can be used for task scheduling and flight control \cite{khalid2023control}, and deep reinforcement learning can optimize integrated sensing and communications (ISAC) \cite{liu2024optimal}.
However, within the context of large-scale connectivity, heterogeneous services, and stringent network performance requirements, these approaches may encounter limitations in terms of data computing efficiency, flight control precision, network performance gains, and implementation costs, decreasing their practicality. Therefore, exploring innovative approaches that address these limitations is crucial for LAE, and EAI is one of the promising solutions. Specifically, EAI is composed of a multi-modal large language model (MMLLM)-based agent and an embodied entity\cite{liu2024aligning}. This makes EAI highly adaptable and capable of interacting with the physical world in a more nuanced and dynamic way.
In this regard, EAI can offer valuable insights into the challenges faced by LAE. The details are as follows:
\begin{enumerate}
\item \emph{EAI for efficient and accurate data processing}. EAI' agents, equipped with powerful MMLLMs, can efficiently process large-scale data streams from various sources such as sensors and user mobile applications. The agents can identify relevant information, filter out noise, and optimize data processing techniques, enabling real-time analysis and decision-making even under constrained computational resources. For instance, in UAV delivery scenarios, MMLLMs can receive command prompts from users and facilitate end-to-end address resolution by using GPS and sensors\footnote{https://arxiv.org/html/2405.01745v1}. By combining MMLLM outputs with point-of-interest information, address resolution can be achieved with high accuracy. As demonstrated in recent studies, this capability can achieve over 90\% accuracy in both offline and online experiments \cite{luo2024language}. 
\item \emph{EAI for robust task scheduling and secure flying control}. EAI can realize task scheduling flight control in LAE by integrating sensory feedback, real-time adaptation, and collaborative decision-making. By learning from interactions users and the environment through embodied entities, the agents can navigate complex airspace, avoid obstacles, and coordinate with other UAVs in a safe and efficient manner. For example, in a safe navigation scenario where a UAV autonomously explores and patrols a complex environment with no prior knowledge of the layout, EAI can notably enhance performance by mitigating challenges associated with sparse rewards during problem solving . In such scenarios, EAI has achieved 88.9 mission success rates for entering into buildings, surpassing the 11.1 success rates achieved by traditional deep reinforcement learning-based approaches\cite{zhao2023agent}. 
\item \emph{EAI for network performance optimization}. EAI can optimize network performance by learning to adapt to dynamic network conditions, predict traffic patterns, and manage resource allocation by incorporating feedback from the physical environment. For example, to simultaneously optimize sensing and communication performance in UAV networks, researchers have formulated a multi-objective optimization problem (MOP) \cite{li2024large}. By decomposing the MOP into sub-problems and utilizing LLM-based agents as black-box search operators, EAI has achieved a joint communication and sensing performance significantly higher than other baselines,
    whose Hypervolume (HV) are 1.194 for the MOP, surpassing the 1.136 HV achieved by the MODEA baseline. 
\end{enumerate}

Therefore, based on above analysis, introducing EAI into LAE can not only solve sensing, communications, computation and control problems faced by traditional approaches, but also offer insights to advance the development of UAV networks. The main contributions of the paper can be summarized as follows:
\begin{enumerate}
\item We highlight the challenges faced by non-AI and disembodied approaches in LAE, which can be better solved by EAI. Besides, we introduce the preliminaries of LAE and EAI-enabled aircrafts and provide a review for above techniques and scenarios, demonstrating the necessities and significances of introducing EAI into LAE.
\item We present an integration of EAI into LAE, mainly from four aspects, sensing, communications, computation and control. In each aspect, we demonstrate the advantages of EAI compared to non-AI or disembodied AI\footnote{Disembodied AI refers to AI that exists purely in software/cyber space, without a physical embodiment, such as ChatGPT, where cognition and physical entities are disentangled} approaches. Furthermore, we highlight the corresponding challenges.
\item We propose a framework of EAI-enabled ISC3 for LAE. We then design an EAI-enabled UAV for an express delivery application a case study. Numerical results have evaluated the effectiveness of the proposed framework.
\end{enumerate}


\section{Preliminaries}

\subsection{Low Altitude Economy}


\begin{figure*}[!htb]
  \centering
  \includegraphics[width=14cm]{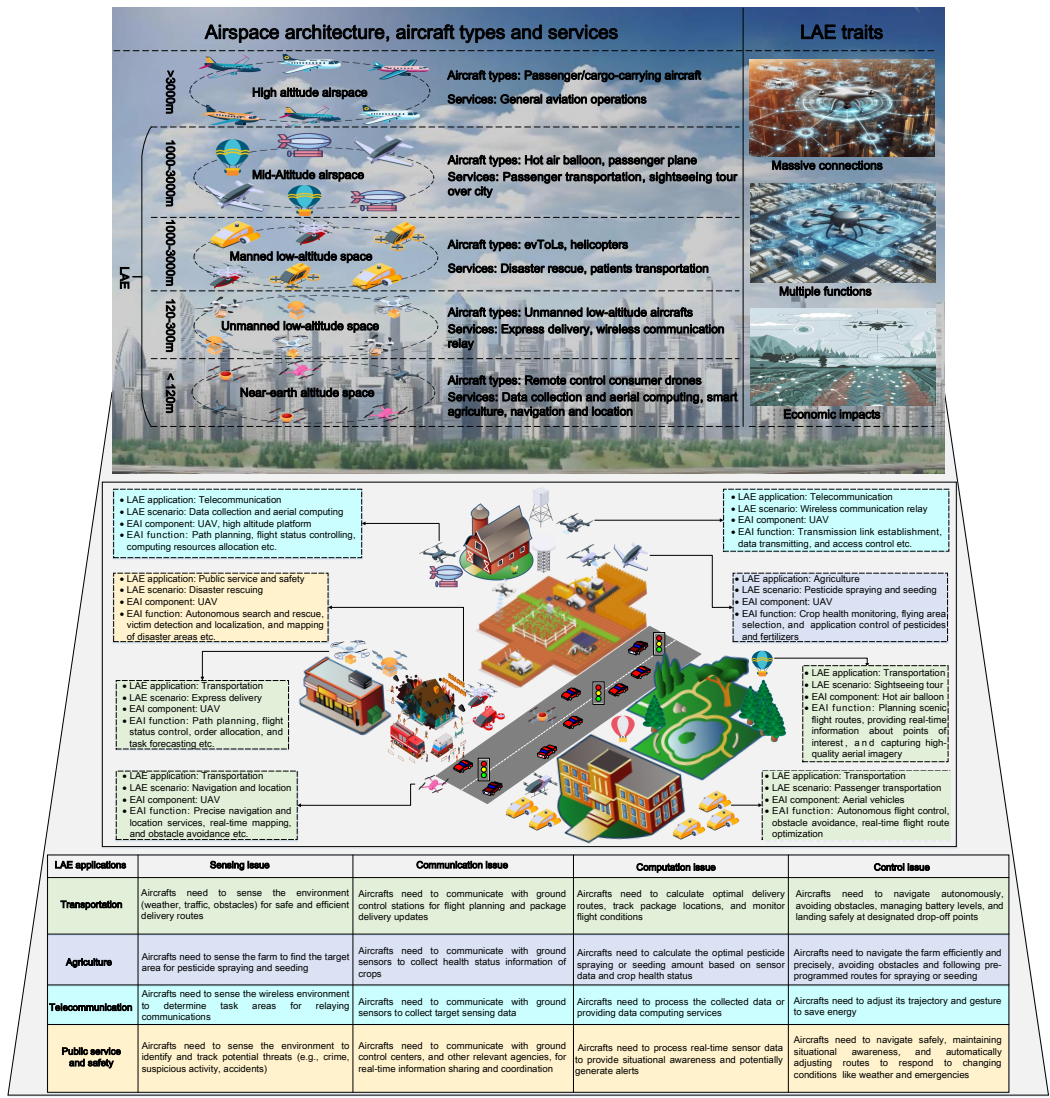}\\
  \caption{Applications of LAE from transportation, agriculture, telecommunication, tourism and and public service and safety fields, and internal relationships between LAE application and the supporting techniques on sensing, communications, computations, and control.}
  \label{APPP}
\end{figure*}
LAE can be a key position in the modern economic system, and its development will further promote the highly-productive and sustainable development of the economy. According to the report by CCID Consulting\footnote{https://www.chinadaily.com.cn/a/202404/09/WS6614977da31082fc043\\c0e2b.html}, LAE in China surged to 69.9 billion dollars in 2023, a 33.8\% year-on-year increase. The market is projected to surpass 1 trillion yuan by 2026. Taking an example, electric vertical take-off and landing (eVTOL) aircraft revenue jumped 77.3\% to 980 million yuan in 2023, with wider commercial use expected in 2024. The eVTOL market is forecast to reach 9.5 billion yuan by 2026, driven by accelerated airworthiness certifications.
Meanwhile, in Germany, the air taxes investment has raised roughly 13 billion dollars since 2019. The eVTOL makers have raised 2.3 billion dollars so far in 2024, compared with 1.5 billion dollars in 2023\footnote{https://money.usnews.com/investing/news/articles/2024-11-25/analysis-liliums-fall-throws-spotlight-on-air-taxi-cash-crunch}.
Furthermore, according to Amazon company\footnote{https://www.aboutamazon.com/news/transportation/amazon-prime-air-drone-delivery-mk30-photos}, the newest Prime Air drone, MK30, will deliver to customers in three US locations as well as cities in Italy and the UK by the end of 2024.
All above and other statistical data corroborate the importance of LAE.
As shown in Fig. \ref{APPP}, LAE
primarily involves the use of manned and unmanned civil aircrafts, encompassing activities such as passenger and cargo transportation, as well as related operations and activities\footnote{https://www.scmp.com/economy/china-economy/article/3275949/whats-buzz-chinas-low-altitude-economy-and-what-propelling-its-development}. Generally, based on the primary functions, LAE's applications can be diverse such as transportation, agriculture, telecommunication, and public services and safety\footnote{https://www.aci-asiapac.aero/media-centre/perspectives/low-altitude-economy-to-fly-high-china-makes-great-strides-in-advanced-air-mobility-development}:
\begin{itemize}
\item LAE for transportation. Transportation is a crucial element of modern economies. LAE can realize express delivery, provide navigation and location services, transport passengers, and disaster relief. Specifically, within the LAE context, UAVs can provide navigation and localization services and deliver merchandise to customers within 120-300 meters. For longer distances, an eVTOL aircraft can be deployed for transportation within 1000-3000 meters, opening new horizons for urban air mobility. Likewise, helicopters can also assist in disaster rescue, search, and transportation of patients.
\item LAE for agriculture. Agriculture is a vital field, providing precious raw materials for food. Other activities include pest control and farm management. In this regard, LAE offers precision farming solutions that can optimize resource use and minimize environmental impact. Specifically, below 120 meters, UAVs can seed, monitor crop health, and perform cost-effective pesticide spraying, contributing to sustainable and productive farming practices.
\item LAE for telecommunication. Telecommunication infrastructure is essential for modern society, but it is often lacking or disrupted in remote areas, during natural disasters, or in emergency situations. In this regard, LAE can provide a flexible and adaptable communication solution based on its valuable data and communication capabilities. Specifically, in telecommunication settings, UAVs and aerial platforms can provide data collection, aerial computing, and communication relay services within 120-300 meters\footnote{https://www.mskyeye.com/what-is-low-altitude-economy/}.
\item LAE for public service and safety. LAE can enhance public service and safety\footnote{https://enterprise.dji.com/public-safety} through surveillance, security, and emergency response. Specifically, traditional ground-based methods may be limited or inaccessible in disaster, while drones equipped with thermal imaging can aid in locating missing persons in challenging terrains.
\end{itemize}


Evidently, within the LAE context, aircrafts will execute complicated tasks to provide highly-efficient, cost-effective, and safe operations. These tasks require four major modules, information perception (sensing), information transmission (communication), data processing for decision-making (computation), and flying and landing (control) \cite{chengleyang2024sensing}. Existing techniques, ISAC, integrated sensing and computation (ISC) and integrated communications and computation (ICC), can only support partial above-mentioned complicated tasks with loose performance requirements. This may lead to failures in several real-time tasks with robust and high precision demands. Therefore, it is necessary and important to realize the ISC3 in LAE. To evaluate the importance the of ISC3, we present relationships between LAE applications and its supporting techniques in Fig. \ref{APPP}.

\subsection{Embodied AI-enabled Aircrafts}

\begin{figure}[!htb]
  \centering
  \includegraphics[width=9cm]{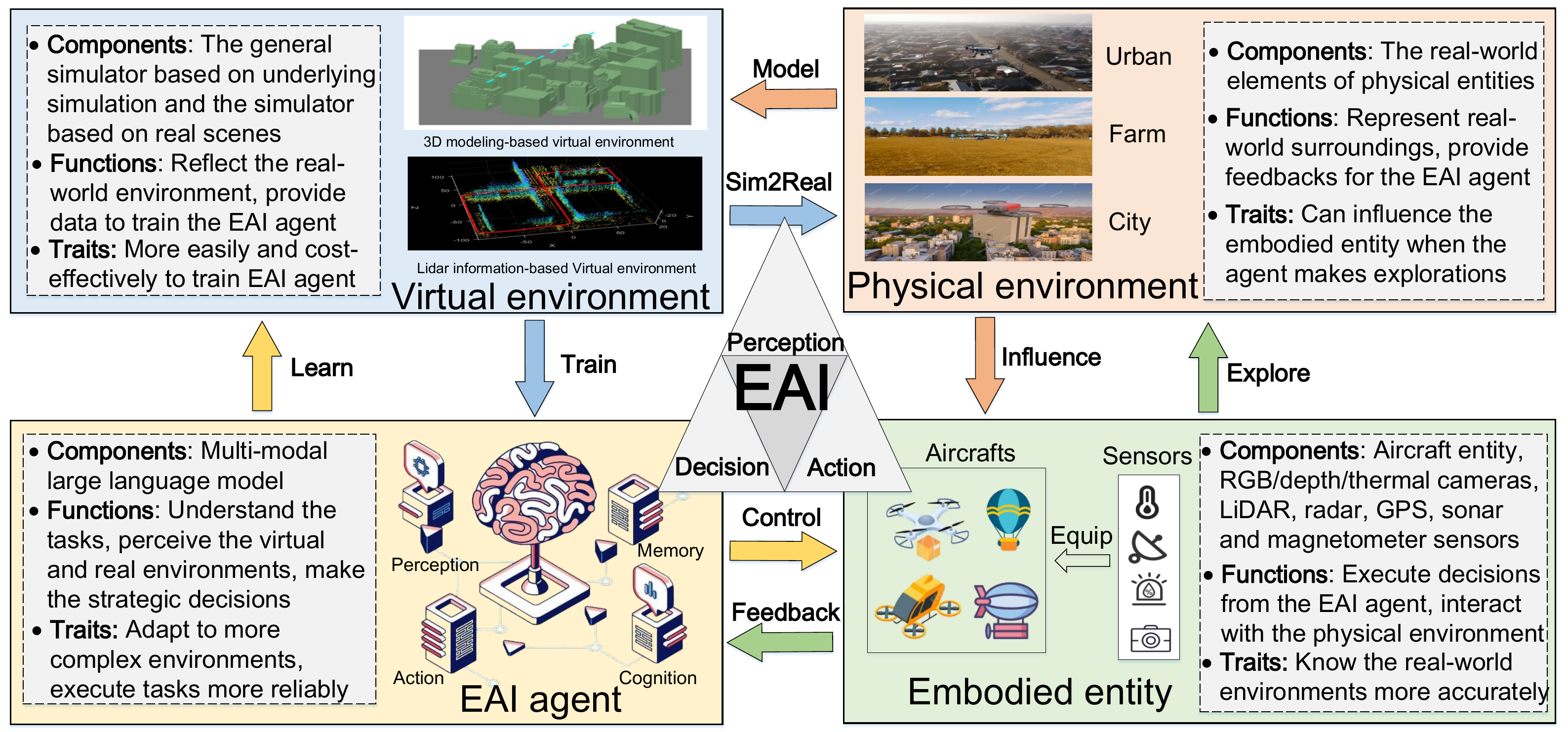}\\
  \caption{Architecture of EAI-enabled aircrafts.}
  \label{EAI222}
\end{figure}

Fig. \ref{EAI222} presents the architecture of EAI-enabled aircrafts. Generally, the EAI-enabled aircraft comprises two main components: an MMLLM-based agent and an embodied entity \cite{liu2024aligning}. The environment consists of both virtual and physical components, interconnected through modeling and sim2real\footnote{Sim2real is a technique in robotics, computer vision, and AI that aims to connect simulated environments and the real world, enabling the efficient training and development of AI agents and robots.} operations. The virtual environment provides training and simulation services for the agent, while the physical environment influences the embodied entity. The embodied entity interacts with the physical environment, gathering data and providing feedback to the agent. The key functions of EAI-enabled aircrafts are organized as follows:
\begin{itemize}
\item Perception. Perception acts as the data collector and processor for the EAI-enabled aircraft, enabling it to understand the external environment and support reliable decision-making and taking actions. This is achieved by fusing multi-modal sensing data from various sensors equipped on the embodied entity. For example, with RGB/depth/thermal cameras, LiDAR, radar, GPS, acoustic sensors and magnetometer sensors, EAI can process above multi-modal sensing data by MMLLM, providing navigation, obstacle avoidance, and target tracking services.
\item Decision. The decision module acts as the command center of EAI-enabled aircraft. Following perception, it executes advanced task planning and reasoning analysis, generating decision-making instructions to control the embodied entity. This process typically leverages the capabilities of the MMLLM. For example, by imputing multi-modal sensing data to MMLLM, EAI can derive strategies for flying path, transmission power allocation, and processing using machine learning models for UAVs with its inference capacity.
\item Action. Action is the execution component of EAI, responsible for receiving decision-making instructions and executing specific actions through the mechanical activity devices of the embodied entity. For the UAV/drone, EAI can control rotor speed, angle, alierons, elevator, and rudder, and also decide the UAV's data transmission/receiving rates, and onboard computers working model (e.g., CPU's working power).
\end{itemize}

Based on above three major functions, EAI can be further applied in UAV swarm systems. It can facilitate better coordination and cooperation between UAVs in a swarm, where UAVs share sensory data, learn from each other¡¯s experiences, and dynamically adjust their roles and tasks based on the needs of the swarm. Furthermore, different from disembodied AI-enabled aircrafts, EAI-enabled aircraft places a strong emphasis on the embodied entity and its interactions with the environment. Through perceiving and actively changing environments with physical interactions, EAI can continuously learn and adapt, leading to improved performance in sensing, communications, computation, and control. \emph{For example, perception in EAI enhances sensing capabilities, decision-making guides communication and computation strategies' making, and action provides feedback for control optimization}. Therefore, leveraging EAI techniques to enhance sensing, communications, computation, and control capabilities is of paramount significance and importance, which can further revolutionize the way we utilize the low-altitude airspace and unlock new possibilities for economic growth and innovation.

\section{Embodied AI for sensing, communications, computation, and control issues in low altitude economic}

\subsection{Research Focus and Solutions}


The framework of EAI-enabled aircraft is composed of an MMLLM and an embodied entity. Specifically, the MMLLM component excels in decision-making and action planning, while the embodied entity facilitates perception through sensor-based data collection.
Therefore, this section explores EAI-enabled advancements in sensing, communications, computation, and control issues that could be further applied in LAE. We examine these issues from the perspectives of:
1) MMLLM-based sensing, communications, and computations. How MMLLM components enhance data acquisition, information transfer, and processing.
2) Embodied entity-based control. How the physical embodiment of the aircraft contributes to robust and adaptable control systems.
This structured approach allows us to comprehensively analyze how EAI empowers LAE systems, highlighting the synergistic interaction between the AI and physical components.

\subsubsection{EAI for Improving Sensing Accuracy}\


Within the LAE context, sensing involves utilizing aircraft to gather data about the environment, primarily through visual images, sensor readings, and other data forms, for applications like mapping, surveillance, environmental monitoring, and precision agriculture. Traditional sensing approaches, relying on manual analysis, basic image processing techniques, or object detection methods like YOLO/FastRCNN, often face limitations in scene interpretation, multi-modal data fusion, and data extraction. However, EAI, with its advanced perception capabilities, offers improvements in scene understanding, data processing, and contextual awareness. For example, a contrastive language-image pre-training model based on LLMs \cite{wang2024knowledge} achieves 73.7\% recognition accuracy for remote sensing images and a quality score of 26.6 for text understanding, demonstrating the potential of EAI in sensing applications. This advancement translates to real-world benefits, such as precision agriculture where EAI-enabled UAVs equipped with multispectral cameras can identify stressed crops, map disease outbreaks, and optimize fertilizer application.

\subsubsection{EAI for Enhancing Communication Rates}\


In the aircraft transmitting data, control commands, and receiving feedback processes, communication in LAE scenarios faces challenges of limited connectivity and unpredictable environments. Traditional approaches, often employing predetermined protocols and optimization-based resource allocation, struggle with bandwidth limitations, interference, and dynamic conditions. EAI, with its advanced decision-making capabilities, offers advantages in resource allocation, interference mitigation, and adaptive learning. For instance, in air-to-ground communication \cite{zhou2024large} utilized LLMs to address transmission power control for ground base stations, achieving performance comparable to deep reinforcement learning (DRL) techniques. This approach eliminates the need for model training and parameter fine-tuning, reaching a transmission cost-oriented reward of 3.8 while maintaining an average data rate of 1.5 Mbps per user. This technique can also be applied to wireless communication relay in LAE, where EAI-enabled UAVs can establish temporary communication links in remote areas, ensuring critical information transmission.

\subsubsection{EAI for Optimizing Computation Resources}\


Computing within the LAE context involves aircraft processing data, executing complex algorithms, and making real-time decisions, often under constraints of limited onboard resources and challenging environmental conditions. Traditional computing approaches, such as centralized or edge computing, they rely on dedicated processors or specialized hardware. This can result in limitations in processing capacity, computational delays, and accuracy. EAI, with its advanced decision-making capabilities, offers advantages in adaptive computing, algorithm optimization, and data processing efficiency. For instance, LLMs have been employed to manage cloud computing resources, achieving significant improvements in CPU utilization, memory usage, network latency, and storage performance \cite{li2024applications}. By optimizing resource allocation, predicting usage patterns, and improving data management, LLMs established a new standard for cloud service management, achieving a 40\% CPU utilization rate and demonstrating read and write speeds of 110 MB/s and 60 MB/s respectively. These advancements are also applicable in aerial computing within LAE, where EAI can assist UAVs in processing collected data, acting as aerial base stations to facilitate emergency data computing.

\subsubsection{EAI for Increasing Control Stability}\

Control within the LAE scenario involves operating aircraft to navigate complex environments, execute tasks accurately, and respond to dynamic conditions while maintaining safety and stability. This encompasses flight control, obstacle avoidance, and adaptive navigation. Traditional control approaches, often based on pre-programmed flight paths or manual control, face limitations in adaptability, manual dependence, and sensitivity to environmental changes. EAI, with its advanced action planning and execution capabilities, offers significant advantages in autonomous flight control, real-time obstacle avoidance, and adaptive navigation. For instance, a framework for navigating multiple dynamic obstacles using an event camera \cite{sanket2019evdodge} utilizes AI design principles that consider hardware limitations and timing/computation constraints. This framework re-arranges the navigation stack as a series of hierarchical competencies, enabling the agent to dynamically avoid obstacles with a 70\% success rate, even in low-light scenarios. This approach can be further applied in transportation within LAE, such as express delivery, where EAI can adapt flight paths in real-time to changing weather conditions or unexpected obstacles, ensuring UAV safety.

\subsection{Lessons Learned}

\begin{table*}[!htb]
\centering
\caption{Research focus and solutions.}
\resizebox{1.0\width}{1.0\height}{
\begin{tabular}{c}\label{SYSTEMMODEL33}
\includegraphics[width=18cm]{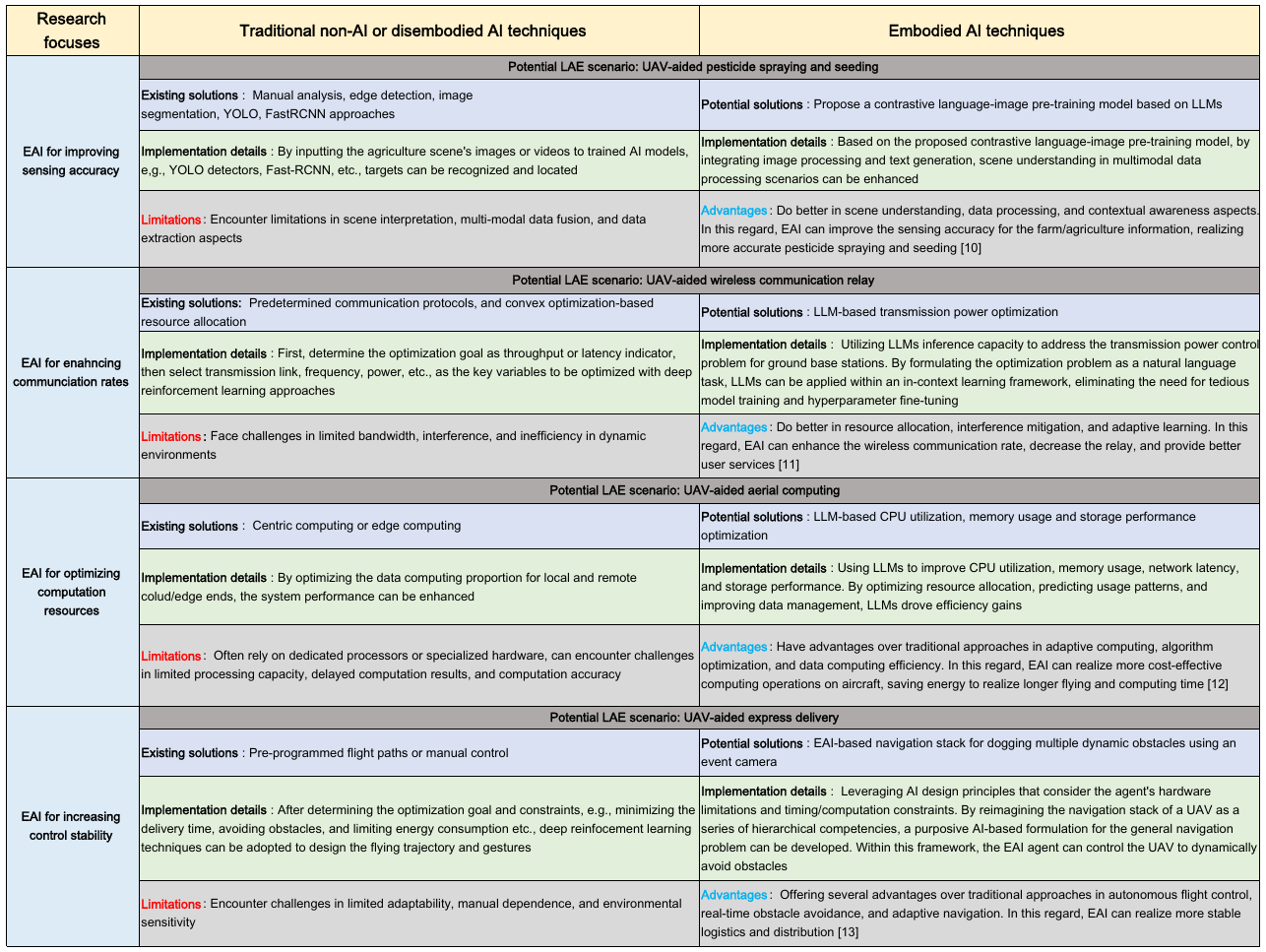}\\
\end{tabular}}
\end{table*}

In summary, EAI can offer significant advancements to LAE across sensing, communications, computing, and control aspects. A comprehensive comparison of traditional approaches and EAI methods in these four areas is presented in Table \ref{SYSTEMMODEL33}. Despite the numerous benefits that EAI can bring to LAE by optimizing sensing, communications, computing, and control performances, its introduction into LAE still faces the following challenges:
\begin{enumerate}
\item Model pre-training requirements: For ease of deployment, EAI models require large amounts of high-quality data to realize pre-training. Ensuring sufficient data availability and addressing data biases are crucial for effective model development and implementation.
\item Malicious and rogue aircrafts concerns: Unauthorized operation of EAI-enabled UAVs poses significant security risks, requiring robust detection and mitigation strategies. This includes measures to prevent unauthorized access and control malicious commands.
\item Law and regulation issues: EAI must follow law and regulation strictly. This imposes new constraints to the ISC3 systems. For example, limitations on flight altitudes, operational areas, data collection practices, and communication protocols may be imposed to ensure safety and privacy.
\end{enumerate}

\section{Embodied AI-enabled ISC3 for LAE}

\subsection{Research Motivation}


Express delivery in LAE refers to utilizing UAVs to transport goods and packages rapidly, especially for short-distance and time-sensitive deliveries. Compared with traditional ground express delivery, UAVs offer advantages including faster delivery times, greater accessibility to remote areas, and reduced traffic congestion. However, large-scale UAV deployment for express delivery presents unique challenges. These include sensing capacity (e.g., integrating data from multiple sensors), communication quality (e.g., real-time interaction with customers and ground stations), computing efficiency (e.g., processing fused multi-source and multi-modal data), and control accuracy (e.g., obstacle avoidance and precise flight path design). To address these challenges, we need to consider a holistic approach integrating sensing, communications, computation, and control. This accounts for base station distribution, weather and terrain conditions, refining UAV flight paths, and implementing effective multi-source data fusion. By integrating these aspects, we can further enhance UAV-based express delivery services, leading to a more efficient and reliable transport system within LAE.

\subsection{Proposed Framework}

\begin{figure*}[!htb]
  \centering
  \includegraphics[width=16cm]{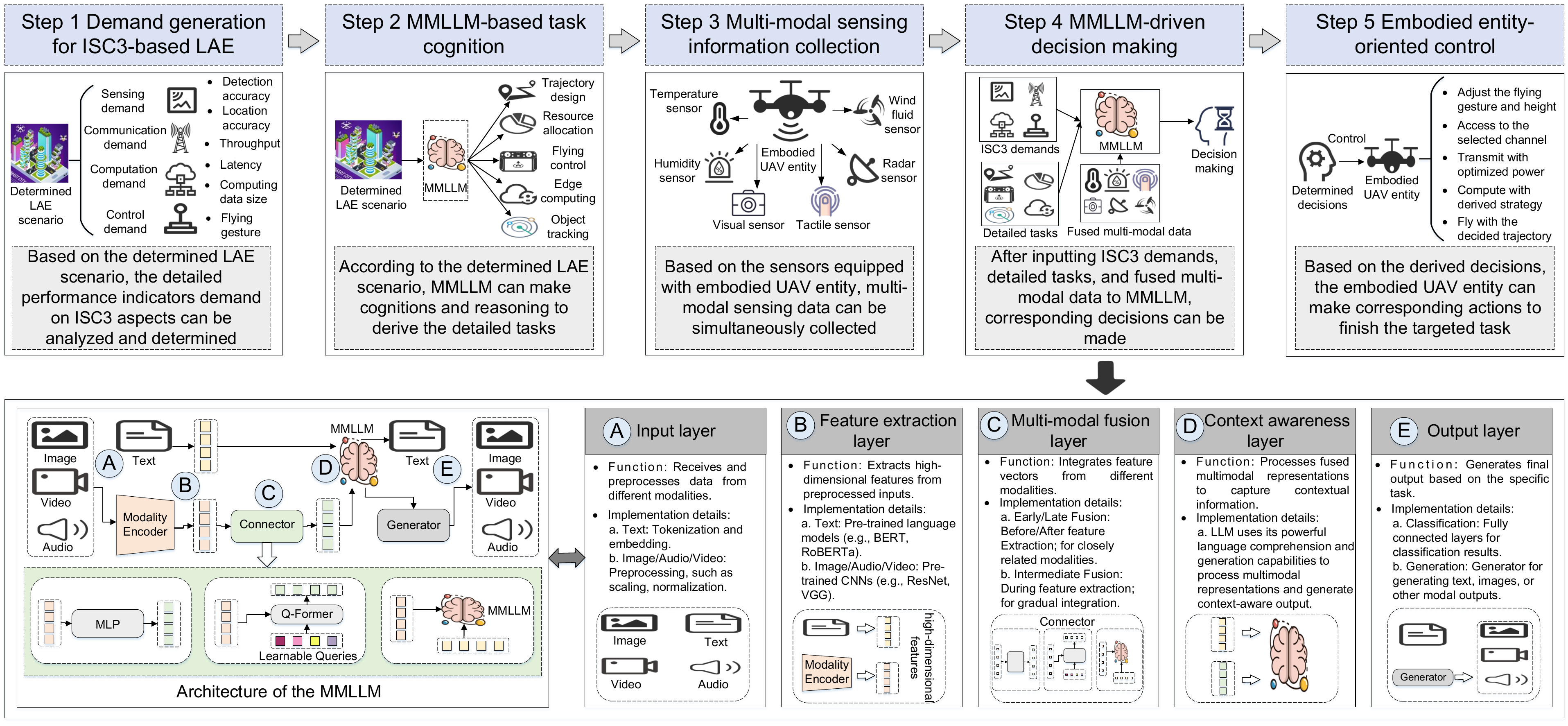}\\
  \caption{Proposed framework of the EAI-enabled ISC3 for LAE. In the MMLLM-driven decision making step, five major layers are included. The input layer receives and preprocesses data from different modalities. The feature extraction layer extracts high-dimensional features from preprocessed inputs. The multi-modal fusion layer integrates feature vectors from different modalities. The context awareness layer processes fused multi-modal representations to capture contextual information. The output layer generates final output based on the specific task.
  }
  \label{SYSTEMMODEL4}
\end{figure*}

With the help of EAI, ISC3 can be realized in LAE. Specifically, as shown in Fig. \ref{SYSTEMMODEL4}, based on the MMLLM\cite{kong2024embodied}, we propose an EAI-enabled ISC3 framework for LAE. This framework can be implemented with the following five major steps:

\textbf{Step 1: Demand generation for ISC3-based LAE.} Drawing upon the specific LAE application scenario, we establish performance indicator demands for the sensing, communications, computation, and control aspects. For instance, this involves defining requirements for object detection and localization accuracy, throughput, latency, computing data rate, and flight trajectory parameters.

\textbf{Step 2: MMLLM-based task cognition.} The MMLLM-based agent, informed by the defined LAE scenario, performs cognitive and reasoning operations to derive detailed tasks. These tasks encompass trajectory design, resource allocation, flight control, edge computing, and object tracking.

\textbf{Step 3: Multi-modal sensing information collection.} UAVs equipped with multiple sensors gather multi-modal and multi-source sensing data. This data is fed into the MMLLM for fusion. For example, temperature, humidity, and wind speed sensors provide weather data, visual cameras capture terrain and transportation details, and radar sensors detect obstacles and distances.

\textbf{Step 4: MMLLM-driven decision making.} After receiving ISC3 demands, detailed tasks, and fused multi-modal data, the MMLLM makes corresponding decisions. Generally, image, video, and audio-based multi-modal data are input to the modality encoder of the MMLLM, while performance indicator demands are input as text\cite{kong2024embodied}. Through reasoning and inference, the MMLLM outputs strategies in text format or generates image, video, or audio-formatted results.
To be specific, based on the five-layer architecture, MMLLM can make ISC3 demand-based decisions with the following implementation details.
\begin{itemize}
\item At first, the input layer is tasked with receiving and preprocessing data from various modalities. For example, for the text data on ISC3 performance demands, the input layer can perform text tokenization and embedding. Image/video/audio-fused multi-modal data is preprocessed by input layer, such as performing scaling, normalization operations.
\item Secondly, the feature extraction layer is responsible for extracting high-dimensional feature representations from the preprocessed inputs. For example, for ISC3 performance demands-based text data, pre-trained language models (e.g., BERT or RoBERTa) are commonly used to extract text features, outputting text feature vectors. Besides, for image/video/audio-fused multi-modal data, pre-trained convolutional neural networks (e.g., ResNet or VGG) are used to extract features, outputting feature vectors.
\item Then, the multi-modal fusion layer integrates feature vectors from different modalities to generate a unified multi-modal representation. Fusion methods can be categorized into early fusion, late fusion, and intermediate fusion. For example, early/late fusion occurs before/after the feature extraction layer and is suitable for scenarios where modalities have strong relationships. Additionally, the intermediate fusion happens during the middle stages of feature extraction and is suitable for scenarios requiring gradual integration of information from different modalities.
\item Next, the context understanding layer further processes the fused multi-modal representation to capture contextual information and interactions between modalities. For example, within an MMLLM, Transformer layers are usually adopted, which employ multi-head self-attention mechanisms to process the fused multi-modal representation, outputting context-aware multi-modal representations.
\item Finally, in the output layer. The specific strategies can be derived based on the ISC3 performance demands. For example, in the text format, MMLLM can classify whether the current command can satisfy the ISC3 performance demands or not. Besides, MMLLM can provide newly aircraft controlling strategies to meet ISC3 performance demands by generating text (e.g., commanding code), images (e.g., flying trajectories), or other modal outputs.
\end{itemize}

\textbf{Step 5: Embodied entity-oriented control.} Based on the obtained strategies, the embodied UAV entity executes corresponding actions to complete the assigned task. This includes adjusting flight trajectory and altitude, selecting communication channels, transmitting at specified power levels, computing using the derived strategy, and following the optimized flight path.

\subsection{Case Study and Numerical Results}

\begin{figure*}[!htb]
  \centering
  \includegraphics[width=14cm]{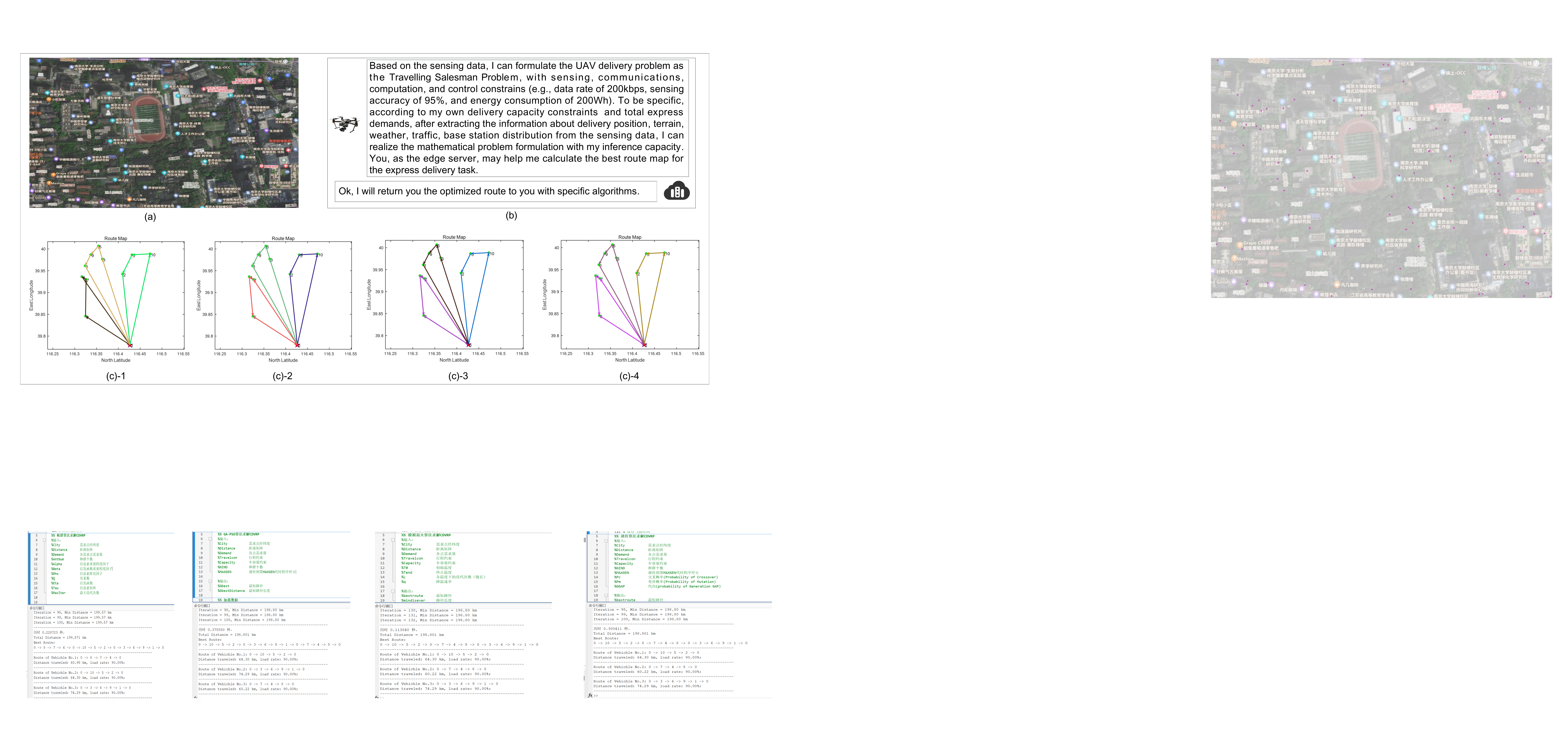}\\
  \caption{Performance evaluation for the proposed framework in EAI-enabled UAV express delivery scenario. (a) Sensing results of the EAI-enabled UAV, which contains information about delivery position, terrain, weather, traffic, base station distribution. (b) Communication and computation processes of the EAI-enabled UAV. (c) Optimized EAI-enabled UAV's route results in control aspect.}
  \label{simu}
\end{figure*}
\textbf{Scenario and goal description.} In this case study, we explore an application of EAI to UAV flight route design for package delivery in LAE. Specifically, an EAI-enabled UAV, at its starting point, can access pre-flight information about weather conditions, traffic patterns, terrain features, and base station distribution within the target delivery area. Recognizing that each package delivery point may have unique demands on qualities of ISC3, and the UAV must maintain optimal communication quality, sensing, computation, and control performance, we can formulate the ISC3 optimization problem. This optimization problem aims to minimize UAV route length while adhering to constraints on communication quality, sensing, computation, and control capabilities. The key variable to optimize is the UAV's flight route. By strategically adjusting the UAV's flight path, we can improve communication quality, maintain sensing effectiveness, ensure efficient computation, and optimize control accuracy, ultimately enhancing the overall efficiency of UAV-based express delivery services.

\textbf{Problem formulation and basic idea.} To achieve efficient express delivery using an EAI-enabled UAV, we formulate two subproblems: (1) optimizing the UAV's trajectory based on predefined ISC3 performance constraints (e.g., e.g., data rate of 200kbps, sensing accuracy of 95\%, and energy consumption of 200Wh), and (2) solving the resulting optimization problem. The basic idea is as below. The UAV, informed by ISC3 requirements, gathers real-time data (delivery locations, terrain, traffic, weather, and base station distribution) within the operational area. This data is processed to identify critical information (obstacles, delivery stations, optimal delivery counts per flight, and current weather conditions). The UAV then models the express delivery route as a Traveling Salesperson Problem (TSP), minimizing flight path length while adhering to constraints like delivery capacity and flight range. Due to energy limitations, the TSP is solved on an edge server using heuristic algorithms\footnote{https://www.crgsoft.com/algorithm-advantages-disadvantages-examples-and-characteristics/}  to ensure real-time performance and stability, avoiding the computational overhead of learning-based or convex optimization methods.
In this regard, above approach highlights the synergistic benefits of EAI: intelligent data acquisition and analysis, efficient decision-making, and offloading computationally intensive tasks to optimize performance.

\textbf{Parameter settings.} The adopted MMLLM is Pixtral 12B\footnote{https://mistral.ai/news/pixtral-12b/}. Besides, the number of express stations are set as 10, the UAV can only carry 20 express one time, and the maximum flying distance of the UAV is 75 km one time. The simulation experiment is executed with Matlab 2023b, on a laptop with Intel(R) Core(TM) i7-11700 CPU @ 1.60GHz, NVIDIA GeForce RTX 3060 GPU, 32.0 GB memory and Windows 11 home operation system.  More detailed simulation settings and code are available at github\footnote{https://github.com/liukewia/Solving-TSP-VRP}.

\textbf{Results evaluation.} Fig. \ref{simu} presents the performance evaluation of the proposed framework in an EAI-enabled UAV express delivery scenario. As shown in Figure \ref{simu}(a), the EAI-enabled UAV effectively gathers necessary sensing data within the task area, including information on delivery locations, terrain, traffic\footnote{https://www.amap.com/}, weather\footnote{http://nmc.cn/publish/forecast} and base station distribution\footnote{https://www.opengps.cn/Data/Cell/Region.aspx}. Then, in Fig. \ref{simu}(b), the EAI-enabled UAV demonstrates communication and computation optimization capabilities when processing the gathered sensing data through MMLLM. For instance, MMLLM can leverage its inference capacity to formulate the TSP \cite{masoud2024exploring}, a mathematical problem, and subsequently offload the route computation task to an edge server. This enables the determination of the optimal route for express delivery. Subsequently, Fig. \ref{simu}(c) illustrates the optimized route of the EAI-enabled UAV in the control aspect.
Specifically, Figs \ref{simu}(c)-1 to \ref{simu}(c)-4 showcase the solutions obtained from applying different algorithms to the formulated TSP (UAV routing problem with travel distance and capacity constraints): ant colony optimization, hybrid particle swarm optimization, simulated annealing, and genetic algorithm, respectively.
The execution times for these algorithms were 0.228723, 0.377555, 0.11304, and 0.300411 seconds, respectively. The shortest route, with a length of 198.801 km, was achieved by the hybrid particle swarm optimization, simulated annealing, and genetic algorithms.
These numerical results validate the feasibility of the proposed framework, where EAI can effectively optimize UAV routes, demonstrating its potential for providing efficient express delivery services within the LAE.

\section{Open Issues}

\subsection{Privacy and Security}
Since LAE involves massive data transmission and processing, privacy and security are very important for aircraft security assurance. In this regard, intrusion detection system (IDS)--based approaches can be adopted to defend against jamming, eavesdropping, and Man-in-the-middle (MITM) attacks.

\subsection{Training Data Augmentation}

The performance of EAI models is related to the quality and quantity of training data. To address potential data scarcity and improve model robustness, data augmentation techniques are crucial. These techniques can include generative AI-based synthetic data generation, image transformations (rotation, scaling, cropping), and noise injection to create a more diverse and representative training dataset.

\subsection{Human-machine Collaboration}

Developing intuitive and safe human-machine interfaces that enable effective collaboration between UAVs and human operators is crucial for ensuring safe and efficient operations. This involves designing user-friendly interfaces that allow operators to monitor UAV activities, provide guidance, and intervene when necessary.

\section{Conclusion}

In this paper, we have introduced EAI into LAE to address concerns in data computing efficiency, flight control precision, and network performance gains. Such an introduction mainly referred to utilizing EAI's perception, decision and action capacities to realize integration of sensing, communications, computation, and control. Specifically, we have first presented preliminaries of LAE and EAI-enabled aircrafts, demonstrating the necessities and significance of introducing EAI into LAE. Then, we have provided some research focuses and solutions of EAI-enabled sensing, communications, computation, and control. Furthermore, we have proposed a framework of EAI-enabled ISC3 for LAE, with an express delivery case study evaluating its effectiveness. At last, we pointed out several future research directions about EAI-enabled LAE.

\bibliography{main}{}
\bibliographystyle{IEEEtran}

\end{document}